\documentclass[11pt]{article}
\usepackage{amsmath}
\usepackage{amscd}
\usepackage{amssymb}
\usepackage{amsfonts}
\usepackage{eufrak}
\usepackage[mathscr]{euscript}
\usepackage{pstricks}
\usepackage{pst-node}
\usepackage[dvips]{pstcol}\textheight=8.5truein
\newcommand{\be}{\begin{equation}}
\newcommand{\ee}{\end{equation}}
\newcommand{\bea}{\begin{eqnarray}}
\newcommand{\eea}{\end{eqnarray}}
\def \Qb{ Q_{\scriptscriptstyle BRS}}
\def \QBb{{\overline {Q}}_{\scriptscriptstyle BRS}}
\def \HT{{\widetilde{\mathcal H}}}
\def \LT{{\widetilde{\mathcal L}}}

\def \QH{{Q_{\scriptscriptstyle H}}}
\def \e{\epsilon}
\def \o{\overline}

\def \t{\theta}
\def \pslsh{\diagup{\hspace*{-.32cm} p}}
\def \tb{\overline{\theta}}
\def \k{\kappa}
\def \kb{\o{\kappa}}
\def \l{\lambda}
\def \ve{\varepsilon}
\def \bc{\o{c}}
\def \w{\omega}
\def \s{\scriptscriptstyle}
\def \v{\varphi}
\def \p{\partial}
\def \QBH{{\overline{Q}}_{\scriptscriptstyle H}}
\def \Qb{ Q_{\scriptscriptstyle BRS}}
\def \QBb{{\overline {Q}}_{\scriptscriptstyle BRS}}
\def \NH{ N_{\scriptscriptstyle H}}
\def \NHB{\overline{N}_{\s H}}

\def \z{\zeta}
\def \zb{\o{\zeta}}

\begin{document}
\baselineskip =15.5pt
\pagestyle{plain}
\setcounter{page}{1}

\title{\huge Classical Mechanics and $\kappa$-Symmetry}
\author{E. Deotto \\ 
Dipartimento di Fisica Teorica, Universit\`a di Trieste, \\
Strada Costiera 11, P.O.Box 586, Trieste, Italy \\ 
and INFN, Sezione di Trieste}
\maketitle

\begin{abstract}
\noindent
In this letter we analyze two local extensions of a  model 
introduced some time ago to obtain a path integral formalism for Classical Mechanics. 
In particular, we show that these
extensions exhibit a nonrelativistic local symmetry
which is very similar to the well known $\kappa$-symmetry introduced in the literature almost 20 years ago.
Differently from the latter, this nonrelativistic local symmetry gives no problem in separating
$1^{\text{st}}$ from $2^{\text{nd}}$-class constraints. 
\end{abstract}
\section {Introduction}
\noindent
The dynamics of relativistic superparticles \cite{Brink} has been deeply analyzed in the last
20 years because of the profound relation between these simple systems and the
more realistic models of supersymmetric field theories and strings. Almost 20 years ago
an important symmetry of the massless supersymmetric particle was discovered by
Siegel \cite{Siegel}. This symmetry, which was also found in superstrings and 
D-branes, allows to gauge away half of the fermionic degrees of
freedom involved in the formalism and has been analyzed in detail in
many following papers \cite{DeAz}-\cite{Moshe}. In particular, a lot of work has been done to 
understand the geometry of the constraints and to solve the problem of quantizing the system.
In fact it is not trivial to quantize the massless superparticle (as well as superstrings and
D-branes) because, due to the presence of the $\kappa$-symmetry, $1^{\text{st}}$-class and
$2^{\text{nd}}$-class constraints cannot be separated covariantly; many attempts have been performed to
solve this problem \cite{Sorokin}\cite{Kallosh}. 

In this letter we continue the analysis (see Ref.\cite{DG}) of the symmetries of a model introduced
some time ago to describe Classical Mechanics in terms of path integrals. This model possesses
a universal {\it global} supersymmetry generated by two charges $\QH$ and $\QBH$. Here we focus on
two other fermionic charges which we call $D_{\s H}$ and $\o{D}_{\s H}$,  which are strictly related 
to $\QH$ and $\QBH$. In fact in superspace $D_{\s H}$ and $\o{D}_{\s H}$ are represented 
by the covariant derivatives associated to the Susy charges mentioned above. 
Following the lines of Ref.\cite{DG} we make these two symmetries ($D_{\s H}$ and $\o{D}_{\s H}$)
local and we note that the new nonrelativistic local Susy we get is very similar 
to the famous $\kappa$-symmetry introduced by Siegel. The main difference with respect to the latter 
becomes manifest after imposing the invariance under local time reparametrization, as one does in Siegel's model.
In fact, in our nonrelativistic framework, there is no difficulty in separating $1^{\text{st}}$-class from 
$2^{\text{nd}}$-class constraints, simply because no $2^{\text{nd}}$-class constraint survives 
after imposing the invariance under local reparametrizations of time. 

There are two simple ways to make local the symmetries $D_{\s H}$ and $\o{D}_{\s H}$ above. The two
models we obtain are two gauge theories which differ in the physical Hilbert space.
We show that one model selects, as physical states, only the distributions built up with the constants of motion only, while
the other is more restrictive and selects only the Gibbs distributions of the canonical
ensemble. 
\section{The $\kappa$-symmetry}
\noindent The model studied by Siegel \cite{Siegel} for the massless relativistic
superparticle is characterized by the following ($1^{\text st}$ order) action:
	\begin{equation}
	\label{1-1}
	S=\int d\tau\left\{p_{\mu}\left[\dot{x}^{\mu}-\frac{i}{2}\left(\zb
	\gamma^{\mu}\dot{\z}-\dot{\zb}\gamma^{\mu}\z\right)\right]-\frac{1}{2}\l
	p^2\right\},
	\end{equation}
\noindent where $x^{\mu}$ are $n$-dimensional space-time coordinates, $\z^{\,\s a}$ and $\zb_{\:\s a}$ are
Dirac spinors and $\l$ is a Lagrange multiplier 
introduced to implement the $p^2=0$ constraint. This action is invariant under the following transformations:
	\begin{align}
	& 
	\text{\bf{$\tau$-reparametrization} (local)} \nonumber \\
	&
	\begin{array}{lll}
	\delta x^{\mu}=\epsilon\dot{x}^{\mu}\,;\hspace{2.8cm} & \delta
	p_{\mu}=\epsilon\dot{p}_{\mu}\,; \hspace{.45cm}& 
	\delta \l=\dot{(\epsilon\l)}\,; \\
	\delta \z=\epsilon\dot{\z}\,; & \delta \zb=\epsilon\dot{\zb}\,; &
	\end{array} \label{1-2}\vspace{.1cm}
	\\
	&
	\text{\bf{Supersymmetry} (global)}\nonumber \\
	&
	\begin{array}{lll}
	\delta x^{\mu}=\displaystyle\frac{i}{2} 
	\left(\o{\ve}\gamma^{\mu}\z-\zb\gamma^{\mu}\ve\right)\,;\hspace{.6cm} 
	& \delta p_{\mu}=0\,; \hspace{.7cm}& \delta \l=0\,; \\
	\delta \z=\varepsilon\,; & \delta \zb=\o{\varepsilon}\,; & \\
	\end{array}\label{1-3}\vspace{.1cm}
	\\
	&
	\text{\bf{$\kappa$-symmetry} (local)}\nonumber\\
	&
	\begin{array}{lll}
	\delta x^{\mu}=\displaystyle\frac{i}{2} 
	\left(\zb\gamma^{\mu}\pslsh\k-\kb\pslsh\gamma^{\mu}\z\right)\,; 
	& \delta p_{\mu}=0\,; & \hspace{.7cm}\delta \l=2i(\dot{\zb}\k-\kb\dot{\z})\,;  \\
	\delta \z=\pslsh\k \,;& \delta \zb=\kb\pslsh\; . &\\
	\end{array}\label{1-4}
	\end{align}
\noindent 
In (\ref{1-2}) the dot means derivation with respect to $\tau$ and $\pslsh$ is 
obviously $p_{\mu}\gamma^{\mu}$. As specified above, $\epsilon$ and $\kappa,\o{\kappa}$ are local
parameters (the first is a commuting scalar, the others are anticommuting spinors) while $\ve$
and $\o{\ve}$ are two global (i.e. they do not depend on the base space $\tau$) 
spinorial parameters. We are particularly interested in the structure of
the third symmetry, which has been deeply analyzed in the literature. Here we want to give a
pedagogical description of the structure of the transformation in phase space, and we want
to highlight the role of the various operators and various commutation structures (Dirac
Brackets) involved. This will turn out to be useful when we will analyze the analog of
the $\kappa$-symmetry in Classical Mechanics.    

First of all we notice that the first and third symmetries above are strictly related. In fact, if we
introduce a mass $m$ in (\ref{1-1}) turning the $p^2=0$ constraint into  $p^2-m^2=0$, we get
 	\begin{equation}
	\label{1-5}
	S_{m}=\int d\tau\left\{p_{\mu}\left[\dot{x}^{\mu}-\frac{i}{2}\left(\zb
	\gamma^{\mu}\dot{\z}-\dot{\zb}\gamma^{\mu}\z\right)\right]-\frac{1}{2}\l
	(p^2-m^2)\right\}.
	\end{equation} 	
\noindent 
$S_m$ is still invariant under (\ref{1-3}) but the other two symmetries
are lost. This is easy to see in
phase space if we apply the Dirac procedure to the actions (\ref{1-1}) and (\ref{1-5}).
Consider first the massive model. The constraints are the following:
	\begin{equation}
	\text{$1^{\text{st}}$-Class} 
	\begin{cases}
	\Pi_{\s \l} = 0 & (a) \\
	p^2-m^2=0 & (b)
	\end{cases} 
	\vspace{.5cm} \hspace{1cm}
	\text{$2^{\text{nd}}$-Class} 
	\begin{cases}
	\Pi_{\s p}^{\mu} = 0 & (c)\\
	(\Pi_{\s x})_{\mu}-p_{\mu} = 0 & (d) \vspace{.1cm}\\ 
	D^{\s a}\equiv(\Pi_{\s\zb})^{\s a}+\displaystyle\frac{i}{2}(\pslsh\,\z)^{\s a} = 0 &(e) \vspace{.15cm}\\
	\o{D}_{\s a}\equiv(\Pi_{\s\z})_{\s a} +\displaystyle\frac{i}{2}(\zb\pslsh)_{\s a} = 0 & (f), 
	\end{cases}\label{1-7}
	\end{equation} 
\noindent
where $\Pi_{(\ldots)}$ are the momenta conjugated\footnote{Here and in the sequel we choose right
derivatives for Grassmannian variables: $\Pi_{\z}:=\frac{\overleftarrow{\p}L}{\p\z}$.} to the variables indicated as $({\s\ldots})$, which
satisfy the following (graded) Poisson Brackets\footnote{In the sequel we shall omit the subcripts $+$ and $-$.}:
	\begin{equation}
	\begin{split}
	& \big[\l,\Pi_{\l}\big]_{-}=1; \hspace{2cm} \big[x^{\mu},p_{\nu}\big]_{-}=\delta^{\mu}_{\nu}; \\
	& \big[\z^{\,\s a},(\Pi_{\z})_{\s b}\big]_{+}=\delta^{\s a}_{\s b}; \hspace{1.2cm}  
	\big[\zb_{\:\s a},(\Pi_{\zb})^{\s b}\big]_{+}=\delta_{\s a}^{\s b}.
	\end{split}
	\end{equation}
\noindent
The first thing to do is to construct the Dirac Brackets associated to the 
$2^{\text{nd}}$-class constraints. If we define the matrix 
	\begin{equation}
	\label{delta}
	\Delta_{ij}=[\phi_i,\phi_j]_{\s PB}
	\end{equation}
where $\phi_k$ are the second class constraints, then the Dirac Brackets between
two generic variables $A,B$ of phase space are defined as:
	\begin{equation}
	\label{1-8}
	[A,B]_{\s DB}=[A,B]_{\s PB}-[A,\phi_i]_{\s PB} 
	(\Delta^{-1})^{ij}[\phi_j,B]_{\s PB}.
	\end{equation}
\noindent
Once we have built the correct structure in phase space, it is not difficult to
realize that the generators of the {\it global} supersymmetry are the following operators:
	\begin{align}
	\label{1-18}
	& Q=\pslsh\,\z; 
	&\o{Q}=\zb\pslsh~\:; 
	\end{align}
\noindent 
which reproduce precisely the transformations (\ref{1-3}) if we define:
	\begin{equation}
	\label{1-18c}
	\delta({\s\ldots})\equiv\big[({\s\ldots}),i\o{\varepsilon}Q-i\o{Q}\varepsilon\big]_{\s DB}.
	\end{equation}
\noindent
Note that the minus sign in the RHS of the previous equation is chosen because of the anticommuting
character of the parameter $\varepsilon$. Moreover we have:
	\begin{equation}
	\label{1-19}
	\big[ Q,\o{Q}\big]_{DB}=i\pslsh,
	\end{equation}
\noindent
which confirms that $Q$ and $\o{Q}$ are two supersymmetry charges. 
Notice that we can induce the same SUSY-transformations through the following operators:
	\begin{align}
	\label{1-20}
	& Q^{\prime}=i\Pi_{\zb}+\frac{1}{2}\pslsh\,\z; 
	& \o{Q}^{\prime}=i\Pi_{\z}+\frac{1}{2}\zb\pslsh; 
	\end{align}
\noindent
which is obvious because $Q\approx Q^{\prime}$ and $\o{Q}\approx \o{Q}^{\prime}$ in the Dirac sense.

Let us now switch to the massless case (\ref{1-1}). The main difference is that we cannot 
repeat all the steps of the previous analysis. In fact the new constraint $p^2=0$ implies that
the matrix $\Delta$ of Eq.(\ref{delta}) is no longer invertible. This is due to the fact
that $\det\Delta\propto\det(\pslsh)=p^{\mu}p_{\mu}=0$. Thus 
the construction of the Dirac Brackets is not as simple as in the massive case. In fact half of 
the constraints in Eqs.(\ref{1-7}-c) and (\ref{1-7}-d) are now $1^{\text{st}}$-class while the other half remains 
$2^{\text{nd}}$-class and the separation of the two sets is not quite easy (see for example
Refs.\cite{Kallosh}). Nevertheless we can list the generators 
of the $\kappa$-transformations of Eq.(\ref{1-4}):
	\begin{align}
	& K=i\pslsh D=i\pslsh\Pi_{\zb}-\frac{1}{2}\pslsh^{\;2}\z; 
	& \o{K}=i\o{D}\pslsh=i\Pi_{\z}\pslsh-\frac{1}{2}\zb\pslsh^{\;2}. \label{1-22}
	\end{align}
\noindent
($K$ and $\o{K}$ generate the transformation (\ref{1-4}) through commutators like those in (\ref{1-18c}).) 
Obviously we should remember that $(K,\o{K})$ are not a set of independent constraints,
as we explained before, because $\pslsh$ is not invertible on the shell of the contraints.
Note that we can write down the form of the generators $K$, $\o{K}$ even if we do not know exactly 
the form of the Dirac Brackets in this particular case. 
We can do that because the $K,\o{K}$ constraints commute (weakly) 
with all the constraints in (\ref{1-7}{\it c}) and (\ref{1-7}{\it d}) and therefore we have 
$[K,({\s\ldots})]_{DB}\approx [K,({\s\ldots})]_{PB}$ (and the same holds for $\o{K}$) whatever are the 
surviving $2^{\text{nd}}$-class constraints determining the Dirac Brackets at hand. 
	
\section {The Functional Approach To Classical Mechanics.}

\noindent
In this section we shall briefly review the path integral approach to Classical
Mechanics which was originally developed in Ref.\cite{Ennio}. The idea
originated from the fact that whenever a theory has an operatorial
formulation, it also possesses a corresponding path integral. Now Classical Mechanics
(CM) does have an operatorial formulation \cite{Koop} and therefore it is reasonable to
look for the corresponding path integral formalism. The strategy to build this
Classical Path Integral (CPI) is simple. In CM we have a $2n$-dimensional phase space ${\cal M}$ 
whose coordinates we denote by $\varphi^{a}$$(a=1,\ldots, 2n)$, i.e.: 
$\varphi^{a}=(q^1,\ldots, q^n; p^1,\ldots, p^n)$, and we indicate with 
$H(\varphi)$ the Hamiltonian of the system. Then, the equations of motion have the form:
	\begin{equation}
	\label{uno}
	{\dot\varphi }^{a}=\omega^{ab}{\partial H\over\partial\varphi ^{b}}\equiv\w^{ab}\p_b H(\v) ~~~~~
	\text{$\w^{ab}=$ symplectic matrix}. 
	\end{equation} 
\noindent 
The classical kernel (i.e. the probability for the system to be in the configuration $\v_f$ at time $t_f$
if it was in the configuration $\v_i$ at time $t_i$) has the following expression:
	\begin{equation}
	\label{due}
	K_{cl}(f|i)=\delta(\varphi^{a}_{f}-\phi^{a}_{cl}(t_{f}|\varphi_{i},t_{i}))
	\end{equation}
\noindent where $\phi^{a}_{cl}(t|\varphi_{i},t_{i})$ is the classical trajectory at time $t$ (that is the
solution of the Hamilton equations) having $\varphi_{i}$ as initial condition at time $t_i$. 
Since $K_{cl}(f|i)$ is a classical probability we can rewrite it as follows:
	\begin{equation}
	\label{tre}
	\begin{array}{rl}
	K_{cl}(f|i) & =\displaystyle\sum_{k_{i}} K_{cl}(f|k_{N-1})K_{cl}(k_{N-1}|k_{N-2})\cdot...\cdot 
	K_{cl}(k_{1}|i)
	\vspace{.2cm} \\
	&=\displaystyle\prod^N_{j=1}\int
	d^{2n}\varphi_{j}~\delta^{(2n)}[\varphi^{a}_{j}-\phi^{a}_{cl}(t_{j} 
	|\varphi_{j-1},t_{j-1})] \vspace{.2cm}\\
	\xrightarrow{N\rightarrow\infty} &=\displaystyle\int{\mathscr
	D}\varphi~\tilde{\delta}[\varphi^{a}(t)-\phi^{a}_{cl}(t)]
	\end{array} 
	\end{equation}
\noindent where in the first equality $k_i$ denotes formally an intermediate configuration $\v_{k_i}$ between 
$\v_{i}$ and $\v_{f}$ and in the last equality the symbol $\tilde{\delta}$ represents a {\it functional} Dirac delta.
The last formula in (\ref{tre}) is already a path integral but we can give it a more familiar form if we rewrite the
Dirac delta as:
	\begin{equation}
	\label{quattro}
	{\tilde\delta}[\varphi^a -\phi^a_{cl}]={\tilde\delta}[{\dot\varphi 
	^{a}-\omega^{ab}
	\partial_{b}H]~\det [\delta^{a}_{b}\partial_{t}-\omega^{ac}\partial_{c}\partial
	_{b}H}]
	\end{equation}
\noindent where we have used the functional analog of the relation
$\delta[f(x)]=\frac{\delta[x-x_i]}{\Bigm|\frac{\partial f}{\partial 
x}\Bigm|_{x_i}}$. 
Next (see Ref.\cite{Ennio} for details) we can exponentiate both terms of the RHS of 
Eq.(\ref{quattro}) via a Lagrange multiplier $\lambda$ (the first term) and a couple of Grassmannian variables
$(c,\o{c})$ (the second term). What we finally get is the following expression:
	\begin{equation}
	\label{cinque}
	K_{cl}(f|i)=\int{\mathscr D}\varphi ^{a}{\mathscr D}\lambda_{a}{\mathscr D}
	c^{a}{\mathscr D}{\o c}_{a}~\exp\biggl[i\int dt~{\widetilde{\cal L}}\biggr]
	\end{equation} 
\noindent where $\widetilde{\cal L}$ is the Lagrangian characterizing the CPI:
	\begin{equation}
	\label{sei}
	{\widetilde{\cal L}}=\lambda_{a}[{\dot\varphi }^{a}-\omega^{ab}\partial_{b}H]+
	i{\o c}_{a}[\delta^{a}_{b}\partial_{t}-\omega^{ac}\partial_{c}\partial_{b}H]
	c^{b},
	\end{equation}
\noindent 
and the $8n$ variables $(\v^a,\l_a,c^a,\bc_a)$
form the new {\it enlarged} phase space which we denote by $\widetilde{\cal M}$.
It is easy to Legendre transform the Lagrangian $\LT$  and obtain the corresponding Hamiltonian:
	\begin{equation}
	\label{sette}
	\widetilde{\cal H}=\lambda_a\omega^{ab}\partial_bH+i\o{c}_a\omega^{ac}
	(\partial_c\partial_bH)c^{b}.
	\end{equation}	 
\noindent From the path integral (\ref{cinque}) we can easily derive \cite{Ennio} the following 
commutator structure:
	\begin{equation}
	\label{otto}
	\big[\v^{a},\l_{b}\big]=i\delta^{a}_{b}~~;~~\big[ 
	c^{a}, {\o c}_{b}\big]=\delta^{a}_{b}. ~~~ \text{(all others are zero)}.
	\end{equation}  
\noindent	
Via these commutators we can realize the $\l$ and ${\o c}$ variables as differential
operators:
	\begin{equation}
	\label{dieci}
	\lambda_{a}=-i{\partial\over\partial\varphi ^{a}};~~~~{\o 
	c}_{a}={\partial\over\partial c^{a}}
	\end{equation}
\noindent and these in turn can be used to construct the operatorial version of the Hamiltonian
(\ref{sette}): 
	\begin{equation}
	\label{undici}
	{\widehat{\widetilde{\cal H}}}\equiv -	i\omega^{ab}\partial_{b}H\frac{\p}{\p\v^a}
	-i\w^{ab}\p_b\p_d H c^{d}\frac{\p}{\p c^a}
	\end{equation} 
\noindent and the corresponding ``Schr\"odinger-type" equation for the probability density $\rho(\v,c;t)$:
	\begin{equation}
	\label{dodici} 
	{\widehat{\widetilde{\cal H}}}\rho(\v,c;t)=i\frac{\p}{\p t}\rho(\v,c;t).
	\end{equation}
\noindent
For a nice interpretation of the geometry of the formalism we refer the reader to Ref.\cite{Geom}.
For our purposes here it is sufficient to say that $\HT$ has a very precise geometrical meaning,
being the {\it Lie derivative} along the Hamiltonian vector field $h\equiv \w^{ab}\p_bH\p_a$.

We end this brief review with some remarks about the symmetries of the Lagrangian (\ref{sei}) and 
the Hamiltonian (\ref{sette}). It is easy to check that they are both invariant under the supersymmetry 
transformations generated by the following operators\footnote{Here we use the same notation as in 
Ref.\cite{Ennio}.}:
	\begin{align}
	\QH & =\Qb-\beta\NH = ic^a\l_a-\beta c^a\p_aH \label{sedici}\\
	\QBH & =\QBb+\beta\NHB= i\o{c}_a\w^{ab}\l_b + \beta\o{c}_a\w^{ab}\p_bH. \label{diciassette}
	\end{align}
\noindent  
($\beta$ is a dimensional parameter). 
It is also not difficult to represent all the formalism developed so far on a suitable 
superspace composed by the time $t$ and two Grassmannian partners $\theta$ and $\o{\theta}$. We refer 
the reader to Ref.\cite{Ennio} for all the details. For our purposes it is sufficient to say that 
we can introduce a classical superfield
	\begin{equation}
	\label{diciannove}
	\Phi^a(t,\t,\tb)=\v^a + \t c^a + \tb\w^{ab}\bc_b +i\tb\t\w^{ab}\l_b.
	\end{equation}
\noindent on which the susy charges (\ref{sedici})(\ref{diciassette}) and the 
Hamiltonian (\ref{sette}) act
as\footnote{According to the formula: $\mathscr{Q}\Phi^a(t,\t,\tb)\equiv[\Phi^a(t,\t,\tb),\e Q]$.}:
	\begin{align}
	&\mathscr{Q}_{\s H} =-\frac{\p}{\p\t}-\beta\tb\frac{\p}{\p t}; 
	&&\o{\mathscr{Q}}_{\s H} =\frac{\p}{\p\tb}+\beta\t\frac{\p}{\p t}.
	&&\widetilde{\mathscr{H}} =i\frac{\p}{\p t};
	\end{align}
\noindent It is also easy to work out the covariant derivatives associated to $\mathscr{Q}_{\s H}$
and $\o{\mathscr{Q}}_{\s H}$:
	\begin{align}
	&\mathscr{D}_{\s H}  =-i\frac{\p}{\p\t}+i\beta\tb\frac{\p}{\p t}; 
	&\o{\mathscr{D}}_{\s H} =i\frac{\p}{\p\tb}-i\beta\t\frac{\p}{\p t}\label{ventitre};
	\end{align}
\noindent which correspond (in $\widetilde{\cal M}$) to the following operators: 
	\begin{align}\label{DD}
	& D_{\s H} =i\Qb+i\beta\NH 
	& \o{D}_{\s H} =i\QBb-i\beta\NHB, 
	\end{align}
\noindent where $\Qb$, $\QBb$, $\NH$ and $\NHB$ are defined in Eqs.(\ref{sedici}) and (\ref{diciassette}).
\section{$\kappa$-symmetry and CPI}

\noindent
In the previous Section we have shown that the formalism of the Classical Path Integral
exhibits a universal {\it global} Supersymmetry.
However, differently from the model of Siegel, it does not possess any local invariance.
If we want to build up a nonrelativistic analog of the model introduced in Section 1,
we first must inject the local $t$-reparametrization invariance into the Lagrangian
(\ref{sei}) by adding the corresponding constraint via a Lagrange multiplier $g$:
 	\begin{equation}
	\label{3-1}
	\LT_1\equiv\LT + g\HT.
	\end{equation}
\noindent
In fact it is easy to see that the previous Lagrangian is {\it locally} invariant under 
   	\begin{equation}
	\label{3-2}
	\begin{cases}
	\delta({\s\ldots})=\big[({\s\ldots}),\e(t)\HT\big] \\
	\delta g=-i\dot{\e}(t). 
	\end{cases}
	\end{equation}
\noindent 
Here and in the sequel $({\s\ldots})$ denotes any one of the variables $(\v^a,\l_b,c^a,\bc_b)$.
Moreover it is easy to check that it remains {\it globally} invariant under the $N=2$
classical Susy of Eqs.(\ref{sedici})(\ref{diciassette}).
Nevertheless, in this simple model no other local symmetry is present. 
If we want to complete the analogy, we must
add (following the lines of Ref.\cite{DG}) two further constraints 
to the Lagrangian (\ref{3-1}) and we get: 
 	\begin{equation} 
	\label{3-16}
	\LT_2\equiv\LT + \xi D_{\s H} + \o{\xi}\o{D}_{\s H} + g\HT.
	\end{equation} 
\noindent
In the previous equation $D_{\s H}$ and $\o{D}_{\s H} $ are the operators introduced 
in Eq.(\ref{DD}).
We want to analyze this model following the same steps we used in Section 1 for the Lagrangian (\ref{1-1}).   

First of all we remember again that, in our non-relativistic case, the analog of the ``$p^2=0$" constraint 
is represented by the term $g\HT$ in (\ref{3-16}) which produces the constraint $\HT=0$. 
Thus, as we did in Eq.(\ref{1-5}), we start our analysis by releasing this constraint in the following
way:  
 	\begin{equation} 
	\label{3-17}
	\LT^{\prime}_2\equiv\LT + \xi D_{\s H} + \o{\xi}\o{D}_{\s H} + g(\HT-\widetilde{E}),
	\end{equation} 
\noindent 
which is the analog of Eq.(\ref{1-5}). It should be remembered that $\widetilde{E}$ is not the
energy of the system, but just a parameter related to the invariance under local time reparametrization:
if $\widetilde{E}=0$ this symmetry is present, while if $\widetilde{E}\neq 0$ this symmetry is lost.

One can immediately work out the constraints:
	\begin{align}
	&\text{$1^{\text{st}}$-Class\hspace{.25cm}}  
	\begin{cases}
	\Pi_{\s \xi} =\Pi_{\s \o{\xi}}=\Pi_{\s g}=0; \\
	\HT-\widetilde{E}=0;
	\end{cases} 
	&\text{$2^{\text{nd}}$-Class\hspace{.25cm}}
	\begin{cases}
	D_{\s H}= 0; \\
	\o{D}_{\s H} = 0. 
	\end{cases}\label{3-18}
	\end{align} 
\noindent
Now we can compare the previous constraints with those in Eq.(\ref{1-7}). Concerning the
$1^{\text{st}}$-class constraints, we notice that $\HT-\widetilde{E}=0$ is the classical analog of the
relativistic mass-shell constraint $p^{\mu}p_{\mu}-m^2=0$. This implies that $\Pi_{\s g}=0$ plays the same
role as $\Pi_{\s\l}=0$ in the relativistic case, while the remaining two constraints ($\Pi_{\s\xi}=0$ and
$\Pi_{\s\o{\xi}}=0$) have no analog in the relativistic case. Consider now the $2^{\text{nd}}$-class constraints.
The first thing to point out is that $D_{\s H}= 0$ and $\o{D}_{\s H} = 0$ are precisely the classical
analogs of $D^{\s a}=0$ and $\o{D}_{\s b}=0$ in the relativistic case. We can say that because $D_{\s H}$ and
$\o{D}_{\s H}$ are related to the classical Susy charges $\QH$ and $\QBH$ in the same way in which $D^{\s a}$
and $\o{D}_{\s b}$ are related to the relativistic Susy charges $Q^{\s a}$ and $\o{Q}_{\s b}$.
In fact it is easy to see that in the relativistic framework $D^{\s a}$ and $\o{D}_{\s b}$ commute with 
$Q^{\s a}$ and $\o{Q}_{\s b}$ and $[D^{\s a},\o{D}_{\s b}]=[Q^{\s a},\o{Q}_{\s b}]=i\pslsh~^{\s a}_{\s b}$
in the same way in which, in the nonrelativistic context, $D_{\s H}$ and $\o{D}_{\s H}$ commute with 
$Q_{\s H}$ and $\o{Q}_{\s H}$ and $[D_{\s H},\o{D}_{\s H}]=[Q_{\s H},\o{Q}_{\s H}]=2i\beta\HT$.    
This is actually the heart of the analogy. 
We start from a model which possesses a universal SUSY generated by $\QH$
and $\QBH$ and we want to check whether it is possible to implement a classical analog of the relativistic
$\kappa$-symmetry of Siegel. Since in the relativistic case the $2^{\text{nd}}$-class constraints are $D^{\s
a}=0$ and $\o{D}_{\s b}=0$, we have modified the CPI-Lagrangian (\ref{sei}) in such a way that the resulting
extension provides as $2^{\text{nd}}$-class constraints the classical analogs of $D^{\s a}$ and $\o{D}_{\s
b}$, that is $D_{\s H}$ and $\o{D}_{\s H}$. This is precisely the model (\ref{3-17}).

If we go on with the same steps as in Section 1 we find that 
the matrix $\Delta_{ij}=[\phi_i,\phi_j]$ has the form:
	\begin{equation} 
	\label{3-20a}
	\Delta =
	\begin{pmatrix}
	0 & 2i\beta\HT \\
	2i\beta\HT & 0
	\end{pmatrix}
	\Longrightarrow
	\Delta^{-1}=
	\begin{pmatrix}
	0 & (2i\beta\HT)^{-1} \\
	(2i\beta\HT)^{-1} & 0
	\end{pmatrix}
	\end{equation} 
\noindent  
and consequently the Dirac Brackets deriving from (\ref{3-18}) are:
 	\begin{equation} 
	\label{3-20b}
	\big[A,B\big]_{\s DB}=\big[A,B\big]- \big[A,\o{D}_{\s H}\big](2i\beta\HT)^{-1}
	\big[D_{\s H},B\big]-\big[A,D_{\s H}\big](2i\beta\HT)^{-1}
	\big[\o{D}_{\s H},B\big].
	\end{equation} 
\noindent
Now that we have the correct structure of our phase space we can proceed with the analogy with the
relativistic case. First of all we can prove that the two supersymmetry charges $\QH$ and $\QBH$ introduced in 
Eqs.(\ref{sedici})(\ref{diciassette}) become weakly equal to the $\Qb$ and $\QBb$ charges:
 	\begin{align} 
	& \QH\approx 2\Qb=2ic^a\l_a; \label{3-21} \\
	& \QBH\approx 2\QBb=2i\bc_a\w^{ab}\l_b; \label{3-22} 
	\end{align}
\noindent and consequently:	
	\begin{align}
	& \big[\Qb,\QBb\big]_{\s DB}=\displaystyle\frac{1}{4}\big[\QH,\QBH\big]_{\s DB}
	=\frac{i\beta}{2}\HT. \label{3-23}
	\end{align}
\noindent
This shows that $\QH$ and $\QBH$ are, more precisely,
the analogs\footnote{This is not in contradiction with what we said few lines above, that 
is that $\QH$ and $\QBH$ are the nonrelativistic analogs of $Q^{\s a}$ and $\o{Q}_{\s b}$. In fact 
it should be remembered that on the shell of the contraints we have $Q\approx Q^{\prime}$, 
$\o{Q}\approx \o{Q}^{\prime}$ (in the relativistic case) and $\QH\approx 2\Qb$,
$\QBH\approx 2\QBb$ (in the nonrelativistic case).} 
of the charges  $Q^{\prime}$ $\o{Q}^{\prime}$ of Eq.(\ref{1-20})
while the $\Qb$ and $\QBb$ charges are analogous 
to the $Q$ and $\o{Q}$ charges of Eq.(\ref{1-18}). 

Consider now the case in which $\widetilde{E}=0$. We get down to the Lagrangian (\ref{3-16}) 
and we see that something happens which is
similar to the mechanism of $\kappa$-symmetry discussed in Section 1. In fact in that case we 
saw that half of the $2^{\text{nd}}$-class constraints became $1^{\text{st}}$-class.
Here, on the other hand, we notice that 
both the  $2^{\text{nd}}$-class constraints $D_{\s H}=\o{D}_{\s H}=0$ become $1^{\text{st}}$-class.
This can be easily seen if one remembers that $\big[ D_{\s H},\o{D}_{\s H}\big]=2i\beta\HT\approx 0$
because now the constraint $\HT-\widetilde{E}=0$ has turned into $\HT=0$.
In other words all the constraints in the model (\ref{3-16}) are gauge constraints and contribute to
restrict the space of the physical states. Therefore we see that in our nonrelativistic framework
there is no difficult in separating $1^{\text{st}}$-class from $2^{\text{nd}}$-class constraints (like 
in the relativistic case). This is simply due to the fact that no  $2^{\text{nd}}$-class constraint remains 
after imposing the constraint $\HT=0$ (which is the classical analog of $p_{\mu}p^{\mu}=0$)\footnote{This 
could be expected somehow, because here we have only two $2^{\text{nd}}$-class constraints and consequently
it cannot happen that only half of these become $1^{\text{st}}$-class, like it happens in the Siegel model.
In fact, if this were the case, we would remain with an odd number (that is 1) of $2^{\text{nd}}$-class 
constraints which is absurd because this number must always be even.}.   

Proceeding with the analogy it is very easy to 
construct the CPI-analogs of $K$ and $\o{K}$ of Eq.(\ref{1-22}), that is 
the generators of the nonrelativistic $\kappa$-symmetry. They are simply: 
  	\begin{align}
	& K_{\s NR}=\HT D_{\s H};
	& \o{K}_{\s NR}=\HT\o{D}_{\s H};	
	\end{align} 
\noindent 
(``$NR$" stands for ``Non Relativistic")
and the local transformations (under which the Lagrangian (\ref{3-16}) is invariant) generated by $K_{\s NR}$ and 
$\o{K}_{\s NR}$ are:
   	\begin{equation}
	\begin{cases}
	\delta({\s\ldots})=\big[({\s\ldots}),\varkappa(t)K_{\s NR}+\o{\varkappa}(t)\o{K}_{\s NR}\big] \\
	\delta\xi=-i\dot{\varkappa}\HT \\
	\delta\o{\xi}=-i\dot{\o{\varkappa}}\HT  \\
	\delta g=2i\beta(\o{\xi}\varkappa+
	\xi\o{\varkappa})\HT. 
	\end{cases}
	\end{equation}
\noindent
It is interesting to determine the physical states
selected by the theory defined by Eq.(\ref{3-16}). Since all the constraints are now $1^{\text{st}}$-class, 
we must impose them strongly on the states as follows:
  	\begin{align}
	&\Pi_{\s \xi} \,\rho(\v,c,\xi,\o{\xi},g) =\Pi_{\s \o{\xi}}\,\rho(\v,c,\xi,\o{\xi},g)
	=\Pi_{\s g}\,\rho(\v,c,\xi,\o{\xi},g)=0; \\
	&D_{\s H}\,\rho(\v,c,\xi,\o{\xi},g)= 0; \\
	&\o{D}_{\s H} \,\rho(\v,c,\xi,\o{\xi},g)= 0; \\
	&\HT\,\rho(\v,c,\xi,\o{\xi},g)=0;
	\end{align} 
\noindent
and it is not difficult to prove that the resulting (normalizable\footnote{Also a state of the form
$\rho(\v,c)\propto \exp[\beta H(\v)]c^1c^2\ldots c^{2n}$ would be admissible, but it is not normalizable in
$\v$.}) physical states have the following form:
 	\begin{equation} 
	\label{3-28}
	\rho(\v,c,\xi,\o{\xi},g)\propto \exp[-\beta H(\v)].
	\end{equation} 
\noindent
This is precisely the {\it Gibbs} distribution characterizing the {\it canonical} ensemble, provided we 
interpret the $\beta$ constant of Eqs.(\ref{sedici})(\ref{diciassette}) as $(k_{\s B}T)^{-1}$, where $T$
plays the role of the temperature at which the system is in equilibrium.
In fact we should remember that up to now the dimensional parameter $\beta$ introduced in 
Eqs.(\ref{sedici}) and (\ref{diciassette}) has not been restricted by any constraint. It is a 
completely free parameter with a dimension of $(\text{\it Energy})^{-1}$ which characterizes
the particular $N=2$ classical supersymmetry. 
The canonical Gibbs state made its appearance earlier in the context of the CPI and precisely in Ref.\cite{Ergo}.
There it was shown that, in the pure CPI model (\ref{sei}), the zero eigenstates of $\HT$ which are also
Susy-invariant are precisely the canonical Gibbs states. In our model instead we have obtained the Gibbs states as
the entire set of physical states associated to the gauge theory described by the Lagrangian (\ref{3-16}).

However the model (\ref{3-16}), though interesting for the peculiar physical subspace it determines, 
is not the nonrelativistic Lagrangian which is closest to the Siegel model. We mean that one should remember
that the Lagrangian (\ref{3-16}) gives rise to a canonical Hamiltonian of the form:
 	\begin{equation} 
	\label{3-29}
	\HT_{2}\equiv\HT -\xi D_{\s H} -\o{\xi}\o{D}_{\s H} - g\HT,
	\end{equation} 
\noindent
but on the other hand we have already checked that the two couples of operators $(\QH,\QBH)$ and 
$(D_{\s H},\o{D}_{\s H})$ close on $\HT$ and not on $\HT_{2}$. Therefore, if we want to 
construct a more precise nonrelativistic analog of the model of Siegel, we should consider a 
slightly modified version of the Lagrangian (\ref{3-16}) which is:
 	\begin{equation} 
	\label{3-30}
	\LT_3\equiv\LT + \dot{\xi} D_{\s H} + \dot{\o{\xi}}\o{D}_{\s H} + g\HT.
	\end{equation} 
\noindent
One can easily check that the Lagrangian (\ref{3-30}) yields, a part from a factor $(1-g)$, the same Hamiltonian as the 
CPI. Therefore we can proceed following the same steps as before: we turn the $\HT=0$ constraint into 
$\HT-\widetilde{E}=0$
 	\begin{equation} 
	\label{3-31}
	\LT^{\prime}_3\equiv\LT + \dot{\xi} D_{\s H} + \dot{\o{\xi}}\o{D}_{\s H} + g(\HT-\widetilde{E})
	\end{equation} 
\noindent
and we find out that the new constraints are:     
	\begin{align}
	&\text{$1^{\text{st}}$-Class\hspace{.25cm}}  
	\begin{cases}
	\Pi_{\s g}=0; \\
	\HT-\widetilde{E}=0;
	\end{cases} 
	&\text{$2^{\text{nd}}$-Class\hspace{.25cm}} 
	\begin{cases}
	\Pi_{\s\xi}+D_{\s H}\equiv D^{\prime}_{\s H}= 0; \\
	\Pi_{\s\o{\xi}}+\o{D}_{\s H}\equiv D^{\prime}_{\s H}= 0.
	\end{cases}\label{3-32}
	\end{align} 
\noindent
Then, it is easy to check that we can repeat all the considerations we did below Eq.(\ref{3-18}), if
we replace $D_{\s H}$ and $\o{D}_{\s H}$ with $D^{\prime}_{\s H}$ and $\o{D}^{\prime}_{\s H}$. As a second
remark, we notice that the two constraints $\Pi_{\s\xi}=\Pi_{\s\o{\xi}}=0$, which had no analog in the
relativistic context, have now disappeared. Moreover, because $\big[D^{\prime}_{\s H},\o{D}^{\prime}_{\s H}\big]=
\big[D_{\s H},\o{D}_{\s H}\big]=2i\beta\HT$, we have also that the Dirac Brackets remain the same as those in
Eq.(\ref{3-20b}), which lead to Eqs.(\ref{3-21})-(\ref{3-23}). Again, when we put $\widetilde{E}=0$, we obtain
that the two $2^{\text{nd}}$-class constraints $D^{\prime}_{\s H}=\o{D}^{\prime}_{\s H}=0$ become both
$1^{\text{st}}$-class, differently from the relativistic case. However, the two models described by the two Lagrangians (\ref{3-16}) and (\ref{3-30})
are not equivalent. There are basically two differences. The first is the new 
form of the nonrelativistic $\kappa$-symmetry which now reads:   
   	\begin{align}
	&\begin{cases}
	\delta({\s\ldots})=\HT\big[({\s\ldots}),\varkappa(t)D_{\s H}+\o{\varkappa}(t)\o{D}_{\s H}\big] 
	\approx
	\big[({\s\ldots}),\varkappa(t)K^{\prime}_{\s
	NR}+\o{\varkappa}(t)\o{K}^{\prime}_{\s NR}\big] \\
	\delta\xi=\big[\xi,\varkappa(t)K^{\prime}_{\s NR}+\o{\varkappa}(t)\o{K}^{\prime}_{\s NR}\big]
	=-i\varkappa\HT \\
	\delta\o{\xi}=\big[\o{\xi},\varkappa(t)K^{\prime}_{\s NR}+\o{\varkappa}(t)\o{K}^{\prime}_{\s NR}\big]
	=-i\o{\varkappa}\HT  \\
	\delta g=2i\beta\HT(\dot{\o{\xi}}\varkappa+
	\dot{\xi}\o{\varkappa}). 
	\end{cases} 
	\end{align}
\noindent where ``$\approx$" is understood in the Dirac sense and
	\begin{align}
	&
	K^{\prime}_{\s NR}\equiv\HT D^{\prime}_{\s H}; \hspace{2cm} \o{K}^{\prime}_{\s NR}\equiv\HT \o{D}^{\prime}_{\s H}.
	\end{align}
\noindent
The second difference, which is the most important, is represented by the two physical spaces associated to the two models 
(\ref{3-16}) and (\ref{3-30}). In fact we have already seen that the physical states associated to the 
first model are the Gibbs distributions $\rho(\v)\propto\exp(-\beta H(\v))$; on the other hand the physical 
states determined by the Lagrangian (\ref{3-30}) must obey the following conditions:
  	\begin{align}
	&\Pi_{\s g}\,\rho(\v,c,\xi,\o{\xi},g)=0~; 
	&D^{\prime}_{\s H}\,\rho(\v,c,\xi,\o{\xi},g) =\big(-i\p_{\xi}+D_{\s H}\big)
	\rho(\v,c,\xi,\o{\xi},g)=0~; \label{3-35}\\
	&\HT\,\rho(\v,c,\xi,\o{\xi},g)=0~;
	&\o{D}^{\prime}_{\s H}\,\rho(\v,c,\xi,\o{\xi},g) =\big(-i\p_{\o{\xi}}+
	\o{D}_{\s H}\big)\rho(\v,c,\xi,\o{\xi},g)=0~.  \label{3-38}
	\end{align} 
\noindent
It is not difficult to realize that the solution of Eqs.(\ref{3-35})-(\ref{3-38}) has the form:
	\begin{equation}
	\rho(\v,c,\xi,\o{\xi},g)\propto\exp\big(-i\xi D_{\s H}-i\o{\xi}\o{D}_{\s H}\big)\tilde{\rho}(\v,c)~,\label{3-39}
	\end{equation}
\noindent where	
	\begin{equation}
	\HT\,\tilde{\rho}(\v,c)=0~, \label{3-40}
	\end{equation}
\noindent
which implies that $\tilde{\rho}(\v,c)$ is a function of constants of motion only. 
Therefore we can say that the physical states associated to the Lagrangian (\ref{3-30}) are isomorphic
to the functions $\tilde{\rho}(\v,c)$ which are annihilated by the Hamiltonian $\HT$ and are
consequently constants of motion. Obviously the Gibbs distributions are a subset of them. This
allows us to claim that the model (\ref{3-30}) is actually more general than 
that characterized by the Lagrangian (\ref{3-16}). More precisely the theory described by 
(\ref{3-30}) is equivalent to that characterized by
the Lagrangian (\ref{3-1}). In fact it is easy to see that the physical Hilbert space associated to the
latter is characterized by the distributions $\tilde{\rho}(\v,c,g)$ obeying to the constraints:
	\begin{align}
	\label{3-41}
	&\displaystyle\frac{\p}{\p g}\tilde{\rho}(\v,c,g)=0; 
	&\HT\,\tilde{\rho}(\v,c,g)=0; 
	\end{align}
\noindent
and the physical space is precisely the same as that in (\ref{3-40}), which is isomorphic to that
determined by Eqs.(\ref{3-35})-(\ref{3-38}).

\section{Conclusions}

\noindent
In this paper we have analyzed two local versions of a model introduced some years ago to 
describe Classical Mechanics in terms of path integrals. In particular, 
we have built two nonrelativistic models which exhibit a universal
local supersymmetry which is very similar to the famous $\kappa$-symmetry introduced
almost 20 years ago by Siegel. Differently from the relativistic case,
in our non relativistic framework the constraint $\HT=0$, which is analogous
to the relativistic $p^2=0$, promotes to $1^{\text{st}}$-class all the $2^{\text{nd}}$-class 
constraints present in the case in which $\HT=\widetilde{E}\neq 0$. 
Consequently there is no difficulty in treating the constraints, differently from what happened in the
relativistic case. In our first model the physical states of the theory turn out to be the Gibbs
distributions characterizing the canonical ensemble, while in the second one the physical 
Hilbert space is formed by all the generic functions of the constants of motion of the 
theory.
\section*{Acknowledgments}

We wish to thank Ennio Gozzi for many fruitful discussions.  
This work has been supported by grants from MURST and INFN of Italy.


\begin{thebibliography}{99}
\bibitem{Brink}
L. Brink, S. Deser, P. Di Vecchia, P. Howe and B. Zumino, Phys.\ Lett.\ B {\bf 64} (1976) 369; 
\bibitem{Siegel}
W.~Siegel, Phys.\ Lett.\ B {\bf 128} (1983) 397;
\bibitem{DeAz}
J.~A.~de Azcarraga and J.~Lukierski, Phys.\ Lett.\ B {\bf 113} (1982) 170;
\bibitem{Sorokin}
D.~P.~Sorokin, V.~I.~Tkach, D.~V.~Volkov and A.~A.~Zheltukhin,
Phys.\ Lett.\ B {\bf 216} (1989) 302; \\
D.~P.~Sorokin, V.~I.~Tkach, D.~V.~Volkov, Mod. Phys. Lett. A {\bf 4} (1989) 901;
\bibitem{Kallosh} 
R. Kallosh, Phys. Rev. D{\bf 56} 3515 (1997); \\
M. Hatsuda and K. Kamimura, Nucl Phys. {\bf B490} 145 (1998);  
\bibitem{Moshe}
M. Moshe and N. Sakai, Phys.\ Rev.\ D{\bf 62}:086004 (2000);
\bibitem{DG}
E.~Deotto and E.~Gozzi,
Int.\ J.\ Mod.\ Phys.\ A {\bf 16} (2001) 2709;
\bibitem{Ennio}
E.~Gozzi, M.~Reuter and W.D.~Thacker, Phys. Rev. D {\bf 40} 3363 (1989);\\
E.~Gozzi, M.~Reuter and W.D.~Thacker, Phys.Rev.D {\bf 46} 757 (1992); 
\bibitem{Koop}
B.O.~Koopman, Proc.~Nat.~Acad.~Sci. USA {\bf 17}, 315 (1931); \\
J.~von Neumann, Ann.Math. {\bf 33},587 (1932); 
\bibitem{Geom}
E.~Gozzi, M.~Reuter, Phys. Lett. B {\bf 240} (1,2) 137 (1990); \\
E.~Gozzi, M.~Regini, Phys.Rev.D {\bf 62} 067702 (2000) (hep-th/9903136); 
\bibitem{Hilbert}
E. Deotto, E. Gozzi and D. Mauro, in preparation.
\bibitem{Planck}
A.A.Abrikosov (jr.), E.Gozzi, Nucl.Phys. B (Proc.Supp.) {\bf vol.88} 369 (2000); 
\bibitem{Vinc}
K.~Sundermeyer, ``{\it Constrained Dynamics}", Springer-Verlag 
Heidelberg, 1982; \\
M.~Hanneaux, C.~Teitelboim, ``{\it Quantization of Gauge Systems}", Princeton
Univ. Press, Princeton, N.J. 1992;
\bibitem{Ergo}
E.~Gozzi, M.~Reuter, Phys.Lett.{\bf 233B} 383 (1989); \\
E.~Gozzi, Prog. Theor. Phys. (Suppl.) {\bf 111}, 115, (1993); \\
E.~Gozzi, M.~Reuter, W.D.~Thacker,  Chaos, Solitons and Fractals 
{\bf 2} 441 (1992); ibid. {\bf 4} 1117 (1994). 

\end{thebibliography}
\end{document}